\documentclass{article}

\usepackage{arxiv}

\usepackage[utf8]{inputenc}
\usepackage[T1]{fontenc}
\usepackage{hyperref}
\usepackage{url}
\usepackage{booktabs}
\usepackage{amsmath}
\usepackage{amsfonts}
\usepackage{nicefrac}
\usepackage{cleveref}
\usepackage{graphicx}
\usepackage{algorithm}
\usepackage{algorithmic}
\usepackage{float}
\usepackage{doi}
\usepackage{array}
\usepackage{hanging}
\usepackage{titlesec}
\usepackage[margin=3cm]{caption}
\usepackage[dvipsnames]{xcolor}
\newcolumntype{P}[1]{>{\centering\arraybackslash}p{#1}}
\graphicspath{ {./images/} }

\title{Quantifying Global Networks of Exchange through the Louvain Method}
\date{}

\titlespacing*{\section}{0pt}{*2}{*0}
\titlespacing*{\subsection}{0pt}{*0.5}{*0.5}

\usepackage{authblk}

\author[1]{%
	\hspace{1mm}Aryan Sharma\thanks{\texttt{\{aryan.sharma0714, jadenmengl\}@gmail.com}; \texttt{\{christina.chu, anna.sisk\}@jhuapl.edu}}%
}
\author[1]{%
	\hspace{1mm}Jaden Li%
}
\author[1]{%
	\hspace{1mm}Christina Chu%
}
\author[1]{%
	\hspace{1mm}Anna Sisk%
}
\affil[1]{National Security Analysis Department, Johns Hopkins University Applied Physics Laboratory}

\renewcommand{\headeright}{}
\renewcommand{\undertitle}{}

\hypersetup{
pdftitle={Quantifying Global Networks of Exchange through the Louvain Method},
pdfsubject={cs.SI},
pdfauthor={Aryan Sharma, Jaden Li, Christina Chu, Anna Sisk},
}

\begin{document}
\maketitle
\vspace{-2em}
\begin{abstract}
Congressional Research Service (CRS) reports provide detailed analyses of major policy issues to members of the US Congress. We extract and analyze data from 2,010 CRS reports written between 1996 and 2024 to quantify inter-country relationships, representing 172 countries as nodes and 4,137 shared interests as edges within a weighted, bidirectional network. Through the Louvain method, we extract non-overlapping communities from our network and identify clusters with shared interests. We then compute the eigenvector centrality of countries to highlight their network influence. The results of this work could enable improvements in sourcing evidence for analytic products and understanding the connectivity of our world.
\end{abstract}
\vspace{0.5em}

\section{Introduction}
Foreign policy is difficult to manage due to the complex and interwoven relationships countries have with one another. In the US government system, the Department of State largely controls foreign policy decisions, while Congressional committees oversee much of the work \cite{congress}. Knowledge of how these relationships are being presented to Congress is a key step in understanding the decisions that the US may make with respect to the rest of the world. We gather data from the Congressional Research Service to examine this phenomenon.

The Congressional Research Service is an agency run by the Library of Congress that aims to provide legislators with detailed research into major policy issues. These open-source reports help members of Congress make informed legislative decisions, thereby maintaining strong and up-to-date policies in all areas of government \cite{research}. To analyze these reports, we propose a weighted graph structure that visualizes clusters of countries and quantifies their influence with respect to US affairs. In this structure, we represent countries as nodes and their shared interests as weighted, bidirectional edges.

\subsection{Louvain Clustering}
The Louvain method \cite{louvain} aims to find a configuration of clusters that maximizes the modularity of a graph. Modularity is a scalar defined in the range $[-1,1]$ that measures the difference in density between inter-cluster and intra-cluster connections. It is formally calculated as:

\begin{equation}
Q=\frac{1}{2m}\sum_{ij}\left[A_{ij}-\frac{k_ik_j}{2m}\right]\delta(c_i,c_j)
\end{equation}

where $m$ is half the sum of all edge weights, $A_{ij}$ is the weight of the edge connecting nodes $i$ and $j$, $k_i$ is the sum of the weights of all edges connected to node $i$, $c_i$ is the cluster that node $i$ belongs to, and $\delta$ is the Kronecker delta function shown below.

\begin{equation}
	\delta(c_i,c_j)=
	\begin{cases}
		1 & c_i=c_j\\
		0 & c_i\neq c_j
	\end{cases}
\end{equation}

To find an ideal set of clusters, the algorithm (Algorithm~\ref{driver}) alternates between two phases. Initially, each node is assigned to its own cluster, after which it is greedily moved to the neighboring cluster that yields the highest modularity gain. Each newly formed cluster is then condensed into a single super-node, and this process is repeated until no node movements can increase modularity. Below is pseudocode demonstrating the implementation.
\vspace{1.5mm}
\begin{algorithm}[H]
\caption{The driver function for the Louvain method}\label{driver}
\begin{algorithmic}[1]
	\STATE \textbf{Procedure} \textsc{Louvain}(Graph $G$, Partition $P$)
		\WHILE {\NOT $done$}
			\STATE $P \gets moveNodes(G, P)$ \textcolor{Emerald}{\COMMENT{Phase 1: Move nodes to increase modularity}}
			\STATE $done \gets length(P) == length(getNodes(G))$
			\IF {\NOT $done$}
				\STATE $G \gets reduceClusters(G, P)$ \textcolor{Emerald}{\COMMENT{Phase 2: Reduce each cluster to a single node}}
				\STATE $P \gets singlePartition(G)$ \textcolor{Emerald}{\COMMENT{Place each newly formed node into its own cluster}}
			\ENDIF
		\ENDWHILE
	\STATE \textbf{End Procedure}
\end{algorithmic}
\end{algorithm}
\begin{algorithm}[H]
\caption{Assign each node to a unique cluster}\label{init}
\begin{algorithmic}[1]
	\STATE \textbf{Procedure} \textsc{singlePartition}(Graph $G$)
		\STATE $newPartition \gets$ []
		\FOR {$v$ in $getNodes(G)$}
			\STATE $newPartition$.append([$v$])
		\ENDFOR
		\RETURN $newPartition$
	\STATE \textbf{End Procedure}
\end{algorithmic}
\end{algorithm}
\begin{algorithm}[H]
\caption{Greedily move nodes into clusters that maximize modularity}\label{phase1}
\begin{algorithmic}[1]
	\STATE \textbf{Procedure} \textsc{moveNodes}(Graph $G$, Partition $P$)
		\WHILE {$currentModularity > oldModularity$}
			\STATE $oldModularity \gets currentModularity$
			\FOR {$v$ in $getNodes(G)$}
				\STATE $maxDeltaQ \gets 0$
				\STATE $maxCluster \gets$ null
				\STATE \textcolor{Emerald}{\COMMENT{Try moving $v$ into each neighboring cluster while keeping track of the greatest modularity increase}}
				\FOR {$cluster$ in $getNeighborClusters(v)$}
					\STATE $deltaQ \gets getModularityChange(v, neighborCluster)$
					\IF {$deltaQ > maxDeltaQ$}
						\STATE $maxDeltaQ \gets deltaQ$
						\STATE $maxCluster \gets cluster$
					\ENDIF
				\ENDFOR
				\IF {$maxCluster$ != null}
					\STATE $P \gets$ move $v$ into $maxCluster$
				\ENDIF
			\ENDFOR
			\STATE $currentModularity \gets getModularity(G)$
		\ENDWHILE
	\STATE \textbf{End Procedure}
\end{algorithmic}
\end{algorithm}
\begin{algorithm}[H]
\caption{Condense each cluster into a single node}\label{phase2}
\begin{algorithmic}[1]
	\STATE \textbf{Procedure} \textsc{reduceClusters}(Graph $G$, Partition $P$)
		\STATE $nodes \gets P$
		\STATE $edges \gets$ []
		\FOR {$A,B$ in P}
			\FOR {$(v,u)$ in $getEdges(G)$}
				\IF {$v$ is in $A$ \AND $u$ is in $B$}
					\STATE $edges$.append($(A,B)$) \textcolor{Emerald}{\COMMENT{If two connected nodes are in different clusters, draw an edge between the clusters}}
				\ENDIF
			\ENDFOR
		\ENDFOR
		\RETURN Graph($nodes$, $edges$)
	\STATE \textbf{End Procedure}
\end{algorithmic}
\end{algorithm}
In some variations of the Louvain method, an additional parameter $\gamma$ is introduced as a coefficient on the $\frac{k_ik_j}{2m}$ term to adjust the resolution (i.e., size) of clusters \cite{resolution}. This parameter behaves as a threshold for density, where intra-cluster connections have a density $d \geq \gamma$ while inter-cluster connections have a density $d < \gamma$. It follows that $\gamma$ is proportional to the number of clusters and inversely proportional to the size of clusters. Our approach to searching for an optimal $\gamma$ is discussed in the Methodology.

\subsection{Eigenvector Centrality}
Centrality generally measures the importance of a node based on its location in a network. Various measures of centrality exist, such as degree centrality, which naively computes centrality as the number of edges connected to a node. We use eigenvector centrality due to its recursive calculation; the centrality of nodes is based on the centrality of their neighbors. Its most notable use occurs in Google's PageRank algorithm \cite{pagerank}.

To calculate eigenvector centrality, we build an adjacency matrix $A$, where $A_{i,j}$ represents the edge weight between nodes $i$ and $j$. We then compute the eigenvector $x$ satisfying $Ax=\lambda x$, where $\lambda$ is the largest eigenvalue in $A$. It follows that the $i$-th entry of $x$ represents the centrality of the node associated with the $i$-th row of $A$.

Since the centrality of a node depends on the centralities of its neighbors, nodes with high eigenvector centrality are connected to other nodes with high eigenvector centrality, which we believe to be an accurate and intuitive depiction of global influence.

\section{Methodology}
We compile a dataset of 2,010 reports from the official CRS database, retrieving an average of ten to fifteen reports per country. All selected reports were published between 1996 and 2024, ensuring breadth and relevance. We run keyword extraction algorithms over this dataset to convert our documents into data points.

Each data point is represented as a tuple $(c, t, f)$, where $c \in C$ is a country, $t$ is a topic of interest, and $f \in \mathbb{N}_0$ is the frequency of said interest. Let $g(c, t)$ denote the frequency of topic $t$ for country $c$, and let $T_{c_i}$ denote the set of topics associated with country $c_i$. We calculate the edge weight between two countries $c_a$ and $c_b$ as $\sum_{t \in T_{c_a} \cap T_{c_b}} g(c_a, t) + g(c_b, t)$. These edge weights are then normalized to the range $(0, 1]$ to improve visualization interpretability.

Performing this calculation across all distinct pairs yields a weighted bidirectional network $G$, consisting of $V = 172$ nodes, $E = 519$ edges and a network density of $E\binom{V}{2}^{-1} \approx 0.0347 = 3.47\%$.

We utilize NetworkX \cite{networkx} to cluster and visualize $G$. To limit the number of clusters, we binary search over the resolution parameter to find the smallest possible $\gamma$ yielding at least ten clusters. This process runs in $O(L(v)\log(\frac{1}{\epsilon}))$ for Louvain runtime $L(v)$ and floating-point precision $\epsilon$. We note that while $L(v)$ is empirically quasilinear, its precise complexity is unproven \cite{runtime}. Once clustering is complete, we analyze relationships among subsets of countries.

\section{Results}
Our network analysis categorized countries into distinct clusters based on shared domestic interests and identified the most influential countries using eigenvector centrality. Figure~\ref{graph} below illustrates an example network generated from our analysis, while Table~\ref{allclusters} enumerates the countries assigned to each cluster (note that the cluster order and country order within clusters are arbitrary).
\begin{figure}[H]
\begin{center}
\includegraphics[scale=0.25]{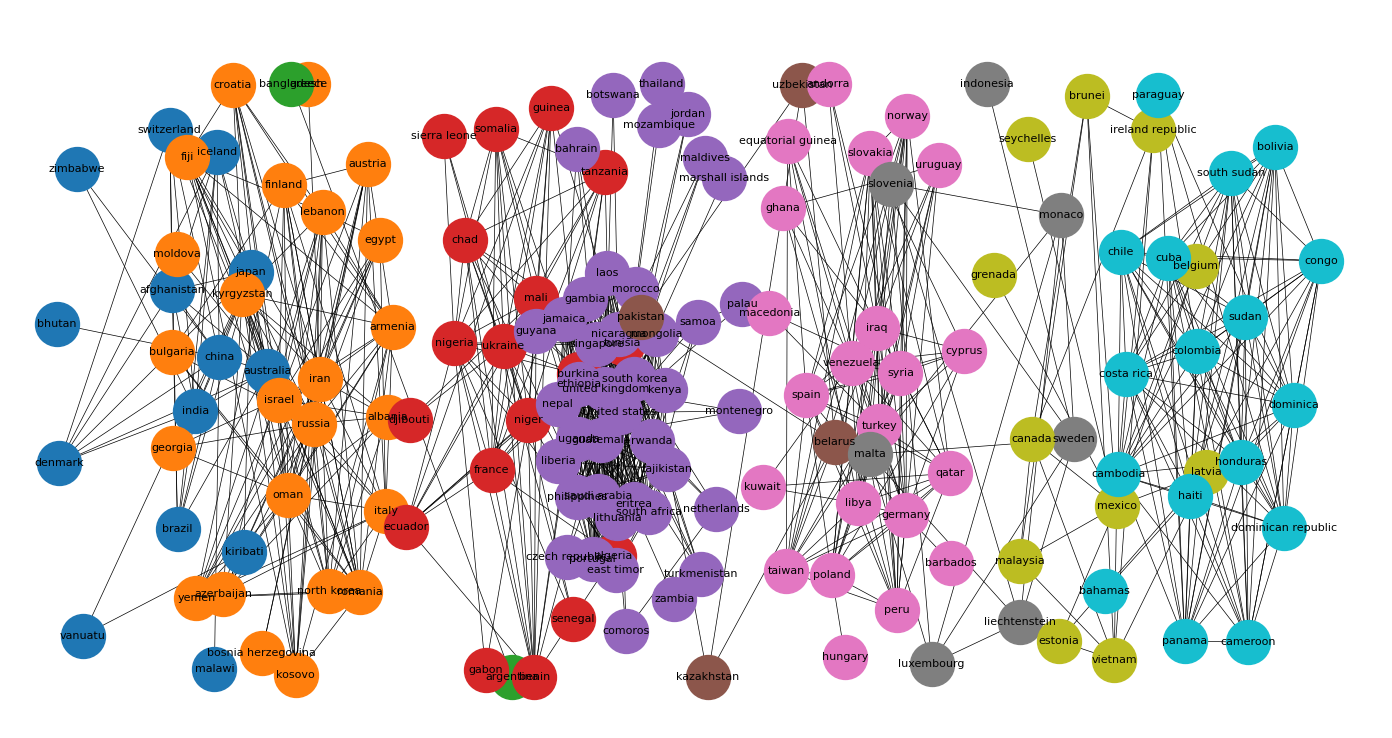}
\caption{A visualization of the complete network with nodes representing countries and edges representing shared interests. The Louvain method identifies ten distinct clusters, each shown in a unique color.}\label{graph}
\end{center}
\end{figure}

\begin{table}[hbt!]
	\caption{Country Groupings by Shared Policy Interests}\label{allclusters}
	\centering
	\newcolumntype{a}{>{\centering\arraybackslash} m{1.5cm} }
	\newcolumntype{b}{>{\centering\arraybackslash} m{13cm} }
	\begin{tabular}{a | b}
		\toprule
		Cluster \# & List of Countries\\
		\midrule
		0 & Afghanistan, Australia, Brazil, China, Bhutan, Denmark, Iceland, India, Japan, Kiribati, Malawi, Switzerland, Vanuatu, Zimbabwe\\
		\hline
		1 & Albania, Armenia, Austria, Azerbaijan, Bosnia Herzegovina, Croatia, Egypt, Fiji, Finland, Bulgaria, Georgia, Greece, Iran, Israel, Italy, Kosovo, Kyrgyzstan, Lebanon, North Korea, Oman, Romania, Russia, Moldova, Yemen\\
		\hline
		2 & Argentina, Bangladesh\\
		\hline
		3 & Algeria, Benin, Burkina, Chad, Djibouti, Ecuador, France, Gabon, Guinea, Mali, Niger, Nigeria, Senegal, Sierra Leone, Somalia, Tanzania, Tunisia, Ukraine\\
		\hline
		4 & Bahrain, Comoros, East Timor, Eritrea, Ethiopia, Gambia, Guatemala, Guyana, Jamaica, Kenya, Laos, Liberia, Lithuania, Maldives, Mongolia, Montenegro, Morocco, Nepal, Netherlands, Nicaragua, Philippines, Portugal, Botswana, Jordan, Rwanda, Samoa, Palau, Saudi Arabia, Singapore, South Africa, South Korea, Tajikistan, Thailand, Turkmenistan, Uganda, United Kingdom, United States, Czech Republic, Marshall Islands, Mozambique, Zambia\\
		\hline
		5 & Belarus, Kazakhstan, Pakistan, Uzbekistan\\
		\hline
		6 & Barbados, Andorra, Cyprus, Equatorial Guinea, Germany, Ghana, Hungary, Iraq, Kuwait, Libya, Norway, Peru, Poland, Qatar, Macedonia, Slovakia, Spain, Syria, Taiwan, Turkey, Uruguay, Venezuela\\
		\hline
		7 & Indonesia, Liechtenstein, Luxembourg, Malta, Monaco, Slovenia, Sweden\\
		\hline
		8 & Brunei, Canada, Estonia, Belgium, Ireland Republic, Latvia, Malaysia, Mexico, Grenada, Seychelles, Vietnam\\
		\hline
		9 & Bolivia, Cambodia, Cameroon, Chile, Colombia, Congo, Costa Rica, Cuba, Dominica, Dominican Republic, Haiti, Bahamas, Honduras, Panama, Paraguay, South Sudan, Sudan\\
		\bottomrule
	\end{tabular}
\end{table}

\begin{figure}[H]
\begin{center}
\includegraphics[scale=0.28]{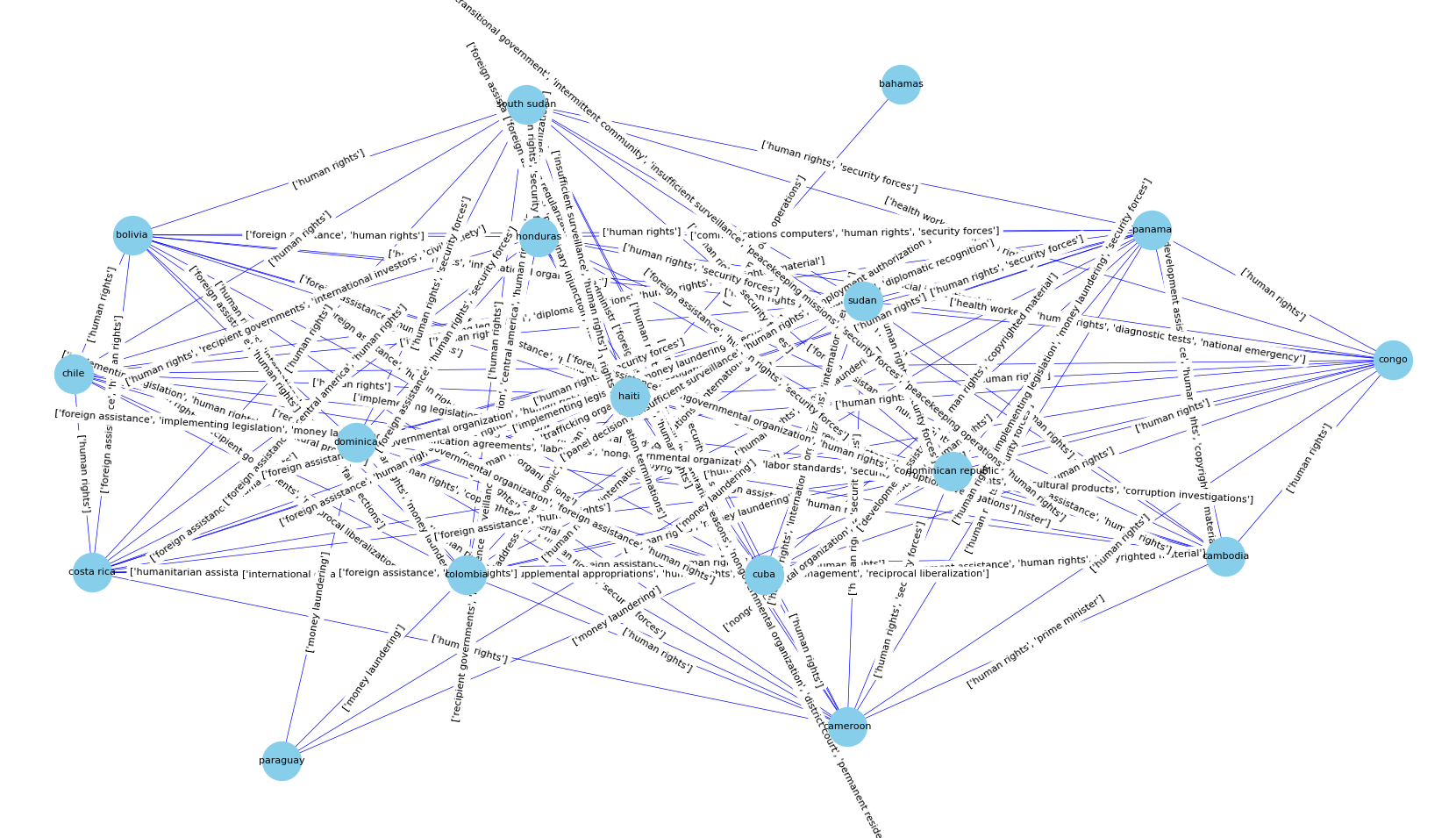}
\caption{A selected cluster containing Colombia, Sudan, and several other countries connected by shared associations with human rights and foreign assistance.}\label{cluster}
\end{center}
\end{figure}

In this run, the countries were organized into ten clusters. We observe that intra-cluster connections are notably denser than inter-cluster connections, verifying the algorithm's effectiveness. Figure~\ref{cluster} shows a magnified view of Cluster \#9 that we arbitrarily select to further analyze.
The edges of this subgraph frequently reference human rights and foreign assistance, likely reflecting the United Nations' involvement with these countries, such as investigations for human rights abuse and deployment of peacekeeping missions. For example, the United Nations Mission in Sudan operated from 2005 to 2011 to stabilize the region, contributing to the formation of South Sudan \cite{un}. Meanwhile, in South America, the US government has investigated Colombia for issues including "arbitrary or unlawful killings" and "serious restrictions on freedom of expression and media freedom" \cite{colombia}. Other clusters are similarly connected by their shared global interests and actions.

\begin{table}[hbt!]
	\caption{Top 10 Countries by Eigenvector Centrality}\label{centrality}
	\centering
	\begin{tabular}{ll}
		\toprule
		Name & Eigenvector Centrality\\
		\midrule
		United States & 0.123\\
		Russia & 0.123\\
		Ukraine & 0.122\\
		China & 0.121\\
		Syria & 0.121\\
		Afghanistan & 0.120\\
		Iraq & 0.120\\
		Mali & 0.119\\
		India & 0.119\\
		Japan & 0.118\\
		\bottomrule
	\end{tabular}
\end{table}

Table~\ref{centrality} lists the ten countries with the highest potential influence in our network as measured by eigenvector centrality. Many notable global superpowers appear on this list, reinforcing the validity of the algorithm's results; it is well known that the U.S. Congress is interested in the activities of these superpowers.

\section{Conclusion}
We apply the Louvain method to a dataset of 2,010 Congressional Research Service reports. By representing 172 countries as nodes and their 4,137 shared interests as bidirectional weighted edges, we extract and visualize recent international relations. Our analysis generates a modular community structure among countries and effectively highlights those with high centrality, enabling analysis of global interactions between countries. These findings improve the quality of evidence in analytic products.

We emphasize that the Louvain algorithm functions as a heuristic, and its inherently greedy nature results in minor variance between different iterations. Increasing the dataset size could capture additional shared interests, resulting in higher network density and likely improving community resolution in turn. We leave this for future work.

\bibliographystyle{ieeetr}
\bibliography{references}

\end{document}